\documentclass{elsart}
\usepackage{graphicx}
\usepackage{hyperref}
\usepackage{natbib}
\usepackage{amsmath}

\journal{Journal of Theoretical Biology}

\begin{document}

\title{Theory of Home Range Estimation from Mark-Recapture Measurements   of Animal
Populations}

\begin{frontmatter}

\author[consortium]{L. Giuggioli},
\ead{giuggiol@unm.edu}
\author[consortium,cab]{G. Abramson},
\ead{abramson@cab.cnea.gov.ar}
\author[consortium]{V. M. Kenkre}
\ead{kenkre@unm.edu}

\address[consortium]{Consortium of the Americas for Interdisciplinary   Science
and Department of Physics, University of New Mexico, Albuquerque, NM 87131,
USA.}

\address[cab]{Centro At\'{o}mico Bariloche, CONICET and Instituto
Balseiro, 8400 San Carlos de Bariloche, R\'{\i}o Negro, Argentina.}

\author{R. R. Parmenter},
\ead{bparmenter@vallescaldera.gov}
\author{T. L. Yates}
\ead{tyates@unm.edu}
\address{Department of Biology, University of New Mexico,
Albuquerque, NM 87131, USA.}

\date{\today}

\begin{abstract}
A theory is provided for the estimation of home ranges of animals from   the
standard mark-recapture technique in which data are collected by   capturing,
tagging and recapturing the animals. The theoretical tool used is the
Fokker-Planck equation, its characteristic quantities being the   diffusion
constant which describes the motion of the animals, and the attractive
potential which addresses their tendency to live near their burrows. The
measurement technique is shown to correspond to the calculation of a   certain
kind of mean square displacement of the animals relevant to the   specific
probing window in space corresponding to the trapping region. The   output of
the theory is a sigmoid curve of the observable mean square  displacement  as a
function of the ratio of distances characteristic of the home range and   the
trapping region, along with an explicit prescription to extract the   home
range form observations. Applications of the theory to rodent movement in
Panama and New Mexico are pointed out.  An analysis is given of the sensitivity
of   our theory to the choice of the confining potential via the use of various
representative cases. A comparison is provided between home range size
inferred from our method and from other procedures employed in the literature.
Consequences of home range overlap are also discussed.
\end{abstract}



\end{frontmatter}

\section{Introduction}

The use of space by animals is the result of a combination of internal factors,
such as the physiology and morphology of the animal, and  external factors such
as the environment. It is well known that mammals, in  order to conduct  their
daily activity, occupy only part of their available environment: the  so-called
home range~\citep{burt43}. A recent study on  the scaling of home range size as
function of animal body mass or metabolic rate~\citep{jetz04} shows that the
home range dimensions are a  compromise between two ingredients: the necessity
for harvesting resources and the detection and response to intrusion. On one
hand, the home range has to be large  enough to meet energy requirements; on
the other, it has to be small enough for  the resident to be protected from
intrusions of same-species foraging neighbours~\citep{buskirk04}. The
importance of learning about home ranges stems not only from the intellectual
need to understand animal movement~\citep{okubo}, but also from the practical
value in the  determination of the size of the home range: it is intimately
related to a variety of ecological phenomena ranging from social organization
to mating behavior and disease transmission~\citep{wolff97,yates02,parmenter83,
abramson2002,abramson2003,vmkpasi,vmkphysicaA,kenkre04}. There is   already a
body of biological literature on home ranges and related animal   movements~%
\citep{okubo,murray}. It is useful to build upon that literature to   provide a
comprehensive mathematical description of the dynamics of ecological  systems.

Given the spatial probability distribution for an individual, the home   range
size has been typically defined as the contour that contains a fixed
percentage (usually 95\%) of the total volume under the distribution~%
\citep{jennrich69,ford79}. Home range size can be estimated from data of
recorded locations over a sufficiently large period of time. These data   can
be obtained through a variety of methods including radio tracking and  live
trapping. Different techniques have been proposed for estimating home   range
size from location data of \emph{single} animals (see e.g. the review
by~\citet{worton87}). The various approaches can be divided into three
categories. In the first, estimates are made using the peripheral  points of
the  location data. In the second, the data are fitted to a pre-assumed
probability distribution. In the third, the probability distribution is
determined  only from the statistical properties of the data (such as  the one
proposed by~\citet{anderson82}). The first method gives the maximum  extent of
the animal's range while the other two methods give a profile of the
probability distribution inside the home range.

In order to obtain reasonable accuracy, all three procedures have the
requirement that the number of locations recorded for each animal be   large~%
\citep{ford79}. Unless unusual efforts are made during the field   sampling,
the number of locations for each individual is typically not very large  (e.g.
\citet{mares80,bergstrom88}). On the other hand, the number of  individuals is
typically quite large. Home range estimation from  location
data of \emph{many} individuals, therefore, avoids such problems~%
\citep{ford79}: the positions are averaged over all the animals   recorded.

The purpose of the present paper is to develop a theoretical model  which gives
a simple prescription for the extraction of home range parameters  from
location data of animal population inside a limited region of space.   Such a
limited window of observation represents, for example, the size of the
trapping array in a mark-recapture experiment, in which an animal is
captured, tagged and then recorded every time it is recaptured~%
\citep{parmenter03}.

In the model we suggest, the motion of each animal is represented by  diffusion
in a confining potential, the latter representing the attraction of the
animal to the home-place, the burrow. The potential has a characteristic width
associated with the size of the home range, which we call $L$. The underlying
equation in our approach is the Fokker-Planck equation for  the probability
distribution for each individual~\citep{okubo,risken}. The stationary solution
of  this Fokker-Planck equation is used to calculate  the infinite-time limit
of  the mean square displacement saturation value  of all the individuals as
function of $L$. Comparison with the measured mean square displacement allows
then the determination of the home range  size, expressed here in units of
length. The home range size has been typically denoted in the literature in
term of  an area. The relation to our description is simply that that area is
given  by the product of $L$ for the two directions. Application of our
procedure to  rodent measurements in Panama and in New Mexico may be found
in~\citep{giuggioli2004,abramson2004} where our model in its simplest form has
been successfully used to  extract not only home range sizes but also diffusion
constants from mark-recaptures of  rodent populations.

The practical output of our present theoretical procedure is a   saturation
curve for the observed mean square displacement as a function of   $L/G$, the
ratio of the home range to $G$, a length that is characteristic of the   size
of the observation window. We predict a sigmoid shape for the saturation
curve. An immediate consequence is that, for the greatest accuracy in  the
measurement of the home range, the observation window should be of the  order
of the  home range. For certain potentials it is possible to write down simple
analytical expressions for the saturation curve. For others the  curve is
obtained  through numerical computation. Our theory also addresses the
distribution of  home ranges according to habitat, equivalently home  range
overlap, a  quantity independently accessible through allometric scaling
arguments~\citep{jetz04}.

The paper is organized as follows. The general problem of calculating   the
average mean square displacement for a population of individuals, each   one
living in its own home range, and observed only inside a spatially   limited
window, is addressed in Sec. \ref{sec-msd}. The sensitivity of the saturation
curve to the choice of the confining   potential is studied in Sec.
\ref{sec-pot} through various representative cases. The case of a non-uniform
distribution of home ranges and considerations for experimentally  determining
the average inter-home distance (related to the home range overlap) is the
subject of Sec. \ref{sec-over}. The comparison between home range size inferred
from our method and the so-called convex polygon  method, usually employed in
the literature, forms Sec. \ref{sec-poly}, and  conclusions are in Sec.
\ref{sec-conc}.

\section{Mean square displacement in a probing window: general
considerations}

\label{sec-msd}

We model the motion of an animal living in its home range through the
Fokker-Planck equation for the probability distribution   $\mathcal{P}(x,t)$
\begin{equation}
\frac{\partial \mathcal{P}(x,t)}{\partial t}=\frac{\partial }{\partial   x}
\left[ \frac{dU(x)}{dx}\mathcal{P}(x,t)\right] +D\frac{\partial   ^{2}\mathcal{%
P}(x,t)}{\partial x^{2}},  \label{FPeq}
\end{equation}
wherein $D$ is the diffusion coefficient of the animal and $U(x)$ is the
potential in which the animal is forced to roam. The potential $U(x)$   is a
representation of the bias or reduced randomness associated with the   walk. A
pure random walk as in a simple diffusive process has $U(x)=0$. When $%
U(x)\neq 0$ we identify its characteristic length with the home  range size
$L$.

The 2-dimensional counterpart of Eq.~(\ref {FPeq}), written in polar
coordinates, is
\begin{eqnarray}
\frac{\partial \mathcal{P}(r,\phi ,t)}{\partial t} &=&\frac{1}{r}\frac{%
\partial }{\partial r}\left[ r\left( \frac{\partial U(r,\phi   )}{\partial r}%
\mathcal{P}(r,\phi ,t)+D\frac{\partial \mathcal{P}(r,\phi ,t)}{\partial   r}%
\right) \right] +  \nonumber \\
&&\frac{1}{r^{2}}\frac{\partial }{\partial r}\left[ \frac{\partial   U(r,\phi )%
}{\partial \phi }\mathcal{P}(r,\phi ,t)+D\frac{\partial   \mathcal{P}(r,\phi
,t)}{\partial \phi }\right].  \label{FPeq2d}
\end{eqnarray}
The description provided by (\ref{FPeq2d}) would be appropriate for a wide
range of animal motion contexts. The third dimension is very rarely required
but can be  easily incorporated. All the essential concepts are, however,
easily represented through the description provided by the 1-dimensional
version. Therefore, we restrict ourselves in the present paper to
Eq.~(\ref{FPeq}). It is straightforward to generalize all  considerations to
higher dimensions if required.

The overall characteristics of the motion can be obtained by  calculating just
the moments of the distribution $\mathcal{P}(x,t)$ rather than the  full
$\mathcal{P}(x,t)$ which entails an integration of the probability. In a
typical experiment a probing window is used, i.e., the animal is   observed
only inside a limited region of space, as in a trapping array in a
mark-recapture experiment. The integration to calculate the moments is  then
performed only over the probing window. The second moment of   $\mathcal{P}%
(x,t)$, i.e., the mean square displacement, is given by
\begin{equation}
\left\langle \Delta x^{2}(t)\right\rangle =\frac{\,\int_{-G/2}^{G/2}dx%
\,(x-x_{0})^{2}\,\mathcal{P}_{x_{0}}(x,t)}{\,\int_{-G/2}^{G/  2}dx\,\mathcal{P}
_{x_{0}}(x,t)},  \label{msd0}
\end{equation}
where $G$ is the dimension of the window and $x_{0}$ is the position of   the
animal at time $t=0$. Because initially each animal can be anywhere   inside $G
$, the numerator and denominator of Eq. (\ref{msd0}) have to be  averaged over
all the possible initial positions inside the window. We then  have,  for the
average,
\begin{equation}
\left\langle \left\langle \Delta x^{2}(t)\right\rangle \right\rangle   =\frac{%
\,\int_{-G/2}^{G/2}dx_{0}\int_{-G/2}^{G/2}dx\,(x- x_{0})^{2}\,\mathcal{P}%
_{x_{c},x_{0}}(x,0)\,\mathcal{P}_{x_{c},x_{0}}(x,t)}{\int_{-G/2}^{G/  2}dx_{0}%
\int_{-G/2}^{G/2}dx\,\mathcal{P}_{x_{c},x_{0}}(x,0)\,\mathcal{P}%
_{x_{c},x_{0}}(x,t)}.  \label{msd1}
\end{equation}
We have introduced here the label $x_{c}$ to represent the burrow   position of
each animal. Equation~(\ref{msd1}) is the contribution to the mean  square
displacement of an animal whose burrow is at $x_{c}$. A further average  over
the distribution of burrow positions is necessary. If this distribution  is
denoted by $\rho (x_{c})$, the observed mean square displacement within  the
window of size $G$ is given by
\begin{equation}
\overline{\Delta x^{2}(t)}=\frac{\int_{-\infty }^{\infty }dx_{c}\rho
(x_{c})\int_{-G/2}^{G/2}dx_{0}\,\int_{-G/2}^{G/2}dx\,(x-  x_{0})^{2}\,\mathcal{P}%
_{x_{c},x_{0}}(x,0)\,\mathcal{P}_{x_{c},x_{0}}(x,t)}{\int_{-\infty   }^{\infty
}dx_{c}\rho   (x_{c})\int_{-G/2}^{G/2}dx_{0}\,\int_{-G/2}^{G/2}dx\,\mathcal{P}%
_{x_{c},x_{0}}(x,0)\,\mathcal{P}_{x_{c},x_{0}}(x,t)}.  \label{msdtd}
\end{equation}

Short-time measurements of $\overline{\Delta x^{2}(t)}$ can be used~%
\citep{giuggioli2004,abramson2004} to obtain the diffusion constant   $D$. In
the present paper we are interested only in the home ranges,   consequently in
the infinite time limit of Eq. (\ref{msdtd}) which requires only the   steady
state solution.

Analytic solutions of the Fokker-Planck equation for all times are  known only
for very few cases of $U(x)$. However, steady state solutions are known  for
any potential explicitly in terms of an  integral~\citep{risken,kuskenkre,pkk}
\begin{equation}
\mathcal{P}_{x_{c},x_{0}}(x,t\rightarrow+\infty )=\frac{e^{   -U(x-x_{c})/D}}{%
\int_{-\infty }^{\infty }dx^{\prime }e^{-U(x^{\prime })/D}}.
\label{steadystateFP}
\end{equation}
Equation (\ref{msdtd}) for $t\rightarrow \infty $ can, thus, be written   as
\begin{equation}
\overline{\Delta x_{ss}^{2}}=\frac{\int_{-\infty }^{\infty }dx_{c}\rho
(x_{c})\int_{-G/2}^{G/2}dx_{0}\,\int_{-G/2}^{G/2}dx\,(x-x_{0})^{2}e^{-  \frac{
U(x_{0}-x_{c})+U(x-x_{c})}{D}}}{\int_{-\infty }^{\infty }dx_{c}\rho
(x_{c})\int_{-G/2}^{G/2}dx_{0}\,\int_{-G/2}^{G/2}dx\,e^{-\frac{%
U(x_{0}-x_{c})+U(x-x_{c})}{D}}},  \label{msdsat}
\end{equation}
and further reexpressed in terms of quantities related to moments of  the
steady state probability density, equivalently of $\exp \left[
-U(x)/D\right] $:
\begin{eqnarray}
\overline{\Delta x_{ss}^{2}} =2\Biggl\{ \int_{-\infty }^{\infty  }dx_{c}\rho
(x_{c})\Biggl[ \left( \int_{-G/2-x_{c}}^{G/2-x_{c}}dx\,e^{-\frac{   U(x)}{D}%
}\right) \left( \int_{-G/2-x_{c}}^{G/2-x_{c}}dxx^{2}e^{-\frac{U(x)}{D}%
}\right) -\Biggr. \Biggr.  \nonumber \\
\left. \left. \left(  \int_{-G/2-x_{c}}^{G/2-x_{c}}dx\,xe^{-\frac{U(x)}{D}%
}\right) ^{2}\right] \right\} \left\{ \int_{-\infty }^{\infty   }dx_{c}\rho
(x_{c})\,\left[ \left(   \int_{-G/2-x_{c}}^{G/2-x_{c}}dx\,e^{-\frac{U(x)}{D}%
}\right) ^{2}\right] \right\} ^{-1}.  \label{msdreduced}
\end{eqnarray}
The expression (\ref{msdsat}) can be reduced further if the burrow
distribution $\rho (x_{c})$ is uniform in space. We obtain
\begin{equation}
\overline{\Delta x_{ss}^{2}}=\frac{G\int_{-G}^{G}dy\,y^{2}g(y)-\left(
\int_{0}^{G}dy\,y^{3}g(y)-\int_{-G}^{0}dy\,y^{3}g(y)\right) }{%
G\int_{-G}^{G}dy\,g(y)-\left(
\int_{0}^{G}dy\,yg(y)-\int_{-G}^{0}dy\,yg(y)\right) },
\label{msdwithg}
\end{equation}
where $g(y)$ is the convolution of $\exp \left[ -U(x)/D\right]$ with   itself:
\begin{equation}
g(y)=\int_{-\infty }^{\infty }dx_{c}e^{-\frac{U\left[ x_{c}\right]   +U\left[
x_{c}-y\right] }{D}}.  \label{gfunction}
\end{equation}
If the potential $U(x)$ remains finite for all finite values of $x$, $%
g(y)=g(-y)$, and it is possible to write a simpler form of   (\ref{msdwithg}):
\begin{equation}
\overline{\Delta x_{ss}^{2}}=\frac{\int_{0}^{G}dy(G-y)y^{2}g(y)}{%
\int_{0}^{G}dy(G-y)g(y)}.  \label{simplermsdsat}
\end{equation}

Clearly, the trapping array of width $G$ can be interpreted as a probe   into
the system whose characteristic width is $L$. The mean square   displacement
depends on the relative magnitude of $L$ and $G$ of the probe. In the   limit
of an infinitely large probe, $\left( \overline{\Delta  x_{ss}^{2}}\right)
^{1/2}$ measures simply the characteristic length of the system. It is  thus
natural to define the home range length $L$ for \emph{arbitrary  potentials} as
the square root of the limit $G\rightarrow \infty $ of Eq. (\ref {msdwithg}),
\begin{equation}
L=\sqrt{\frac{\int_{-\infty }^{+\infty }dy\,y^{2}g(y)}{\int_{-\infty
}^{+\infty }dy\, g(y)}},  \label{hrdef}
\end{equation}
when such a limit exists. If the probe is very small compared to the home range
width, the trapping array will measure a quantity associated with the width of
the  grid. In fact $g(y)$ becomes a constant in the limit $L\rightarrow  \infty
$  and Eq. (\ref{simplermsdsat}) gives $G^{2}/6$ for the mean square
displacement \footnote{While the definition (\ref{hrdef}) of the home range $L$
is the most natural, alternate definitions are possible as used,  e.g., in
\citep{abramson2004}.}.

In a recent article, one of the present authors~\citep{kenkre05} has   given an
alternate formulation in terms of the Fourier transform of the steady  state
probability distribution (\ref{steadystateFP}). It has been shown there  that
the mean square displacement is given simply in terms of Fourier-space
integrals of the product of the square of the sinc function (which carries
information about the probe) with respectively the  square, and the derivative
of the square, of the transform of the steady state probabilities (which carry
information about the home ranges). Alternative expressions equivalent to
Eq.~(\ref{msdwithg}) and (\ref{hrdef}), respectively, are given in
\citep{kenkre05} as
\begin{equation}
\overline{\Delta x_{ss}^{2}}=-\frac{\int_{-\infty }^{\infty }dk\frac{%
\partial ^{2}\widehat{P}^{2}(k)}{\partial k^{2}}\frac{\left[ 1-\cos
(Gk)\right] }{k^{2}}}{\int_{-\infty }^{\infty   }dk\widehat{P}^{2}(k)\frac{%
\left[ 1-\cos (Gk)\right] }{k^{2}}}  \label{fouriertmsd}
\end{equation}
and
\begin{equation}
L^{2}=-2\left. \frac{\frac{\partial ^{2}\widehat{P}(k)}{\partial   k^{2}}}{%
\widehat{P}(k)}\right| _{k=0}=2\frac{\int_{0}^{+\infty   }dy\,y^{2}e^{-\frac{U(y)%
}{D}}}{\int_{0}^{+\infty }dy\,e^{-\frac{U(y)}{D}}},  \label{fouriertL}
\end{equation}
where $\widehat{P}(k)$ is the Fourier transform of $\exp \left[
-U(x)/D\right] $.

Clearly, a dimensionless quantity of crucial importance to the analysis  is the
ratio $\zeta$ of the home range to the observational probe length $G$:
\begin{equation}
\zeta=L/G.
\label{zetadef}
\end{equation}
In the next section we study the functional dependence of the  saturation curve
on this quantity $\zeta$ with attention to the effects of the details  of the
confining potential, assuming that $\rho(x_{c})$ is a constant.

\section{Dependence on the details of the confining potential}

\label{sec-pot}

The precise shape of the confining potential $U(x)$ obviously depends  on the
detail of animal movement, such as habitat and distance between neighbours.
Since this detail is largely unavailable, it is important to determine the
sensitivity of the deduced value of the home range size $L$ to the   choice of
$U(x)$. It is clear that, when plotted as a function of $\zeta$ (see  Eq.
(\ref{zetadef})), the mean square displacement $\overline{\Delta x_{ss}^{2}}$
for each   potential starts out as $L^{2}$ when $\zeta\rightarrow 0$ and
saturates to  $G^{2}/6$ when $\zeta\rightarrow \infty$.

\begin{figure}[h]
\centering  \resizebox{\columnwidth}{!}{\includegraphics{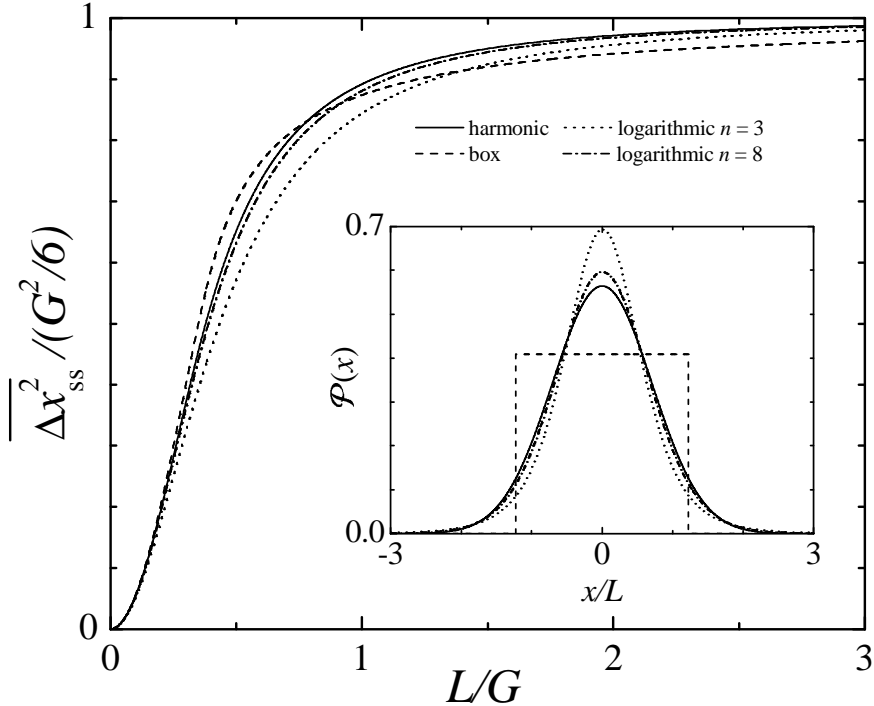}}
\caption{Mean square displacement at saturation for four different  potentials:
a box potential (dashed line), a harmonic potential (solid line), and  two
logarithmic potentials (see text for definition) one with $n=3$ (dotted line)
and  the other with $n=8$ (dash-dotted line). The inset shows the corresponding
stationary distributions $\mathcal P(x)$ as obtained from Eq.
(\ref{steadystateFP}) by putting $x_{c}=0$.}
\label{msdcomparison}
\end{figure}

Extensive studies we have carried out with different potentials have  made it
clear that the curvature at the bottom and the steepness with which $U(x)$
becomes  infinite both play a role in shaping the saturation curve of
$\overline{\Delta  x_{ss}^{2}}$. More precisely, the rise of $\overline{\Delta
x_{ss}^{2}}$ is controlled by the steepness of the  potential when $x/L\gg 1$:
the steeper the rise to infinity  of the potential, the smaller the value of
$\zeta$ for which the saturation curve grows faster than the $L^{2}$ dependence
at $L=0$.  In addition, the curvature of $U(x)$ for $x/L\ll 1$  determines the
way $\overline{\Delta x_{ss}^{2}}$ approaches the value $G^{2}/6$: the  larger
the curvature, the slower it approaches the asymptote $G^{2}/6$. In Fig.
\ref{msdcomparison} we show this dependence by comparing four  characteristic
potentials: a box potential, a harmonic potential and two types of  logarithmic
potentials.

\subsection{Box and harmonic potentials}

The box potential has the steepest rise among the four curves in Fig.
\ref{msdcomparison}, since it diverges at a finite distance. Observe also
that it is the slowest to reach $G^{2}/6$, being the one with the smallest
curvature (zero) close to the origin. Such a potential has been considered in
our previous work~\citep{giuggioli2004,abramson2004} for extracting home  range
sizes from mark-recapture data for two different rodent populations. In those
investigations the saturation curve was numerically simulated.  However, it is
possible to calculate analytically all the integrals in Eq. (\ref {msdreduced})
by selecting judiciously the limit of integration for the variable  $x_{c}$ as
function of the relative dimension of the box width and the probe  width. The
resulting  expression in terms of $\zeta$ is given by
\begin{equation}
\frac{\overline{\Delta x_{ss}^{2}}}{G^{2}/6}=\left\{
\begin{array}{ll}
\frac{18\zeta ^{2}}{5}\frac{(5-3\sqrt{6}\zeta )}{(3-\sqrt{6}\zeta )}, &
\,\,\,\,\,\,\mbox{for }\zeta <\sqrt{6}, \\
\frac{3}{5}\frac{(3-5\sqrt{6}\zeta )}{(1-3\sqrt{6}\zeta )}, &   \,\,\,\,\,\,%
\mbox{for }\zeta >\sqrt{6}.
\end{array}
\right.   \label{msdbox}
\end{equation}
The limiting behaviour is that $\overline{\Delta  x_{ss}^{2}}\simeq
L^{2}$ for small $\zeta$ and $\overline{\Delta x_{ss}^{2}}\simeq  G^{2}/6-4/(15%
\sqrt{6}\zeta)$ for large $\zeta$.

The harmonic potential has been used in one of our previous studies~%
\citep{abramson2004} for extracting the home range size for the deer mouse,
\emph{Peromyscus maniculatus}, in  New Mexico. The shape of the potential is
given by $U(x)=D(x/L)^{2}$. Since the domain where $U(x)$ is not zero extends
over the entire real axis, it is convenient to calculate the convolution
(\ref{gfunction}) which gives $g(y)=L \sqrt{\pi /2}\exp \left[
-y^{2}/(2L ^{2})\right] $. Integrating Eq. (\ref{simplermsdsat})  gives
the mean square displacement at saturation as
\begin{equation}
\frac{\overline{\Delta x_{ss}^{2}}}{G^{2}/6}=6\zeta ^{2}\left\{   1+\frac{%
\sinh \left[ \left(1/2\zeta \right)^{2}\right] }{\sinh \left[ \left(   1/2\zeta
\right)^{2}\right] -\frac{1}{2\zeta }\sqrt{\frac{\pi }{2}}e^{\left(   1/2\zeta
\right) ^{2}}\mathop{\rm erf}\nolimits\left( \frac{1}{\sqrt{2}\zeta }%
\right) }\right\} ,  \label{msdparabola}
\end{equation}
with $\overline{\Delta x_{ss}^{2}}\simeq L^{2}$ when $\zeta\ll 1$ and $%
\overline{\Delta x_{ss}^{2}}\simeq G^{2}/6\left[ 1- \left(7/120\zeta^{2}\right)
\right] $ when $\zeta\gg 1$ . The  intersection in  Fig. \ref{msdcomparison}
between the harmonic potential curve (solid) and  the box potential curve
(dashed) becomes evident here given that the former  approaches $G^{2}/6$
linearly while the latter approaches it  quadratically as  $\zeta \rightarrow
\infty $.

\subsection{Logarithmic potentials}

As an example of $U(x)$, whose corresponding steady state distribution given by
(\ref{steadystateFP}) decays to zero slower (algebraically) than in the box or
the harmonic case, we consider a family of   potentials of the form
\begin{equation}
U(x)=D\ln \left\{ 1+\frac{x^{2}}{\left(\kappa  _{n}L\right)^{2}}\right\}^{n},
\label{logpot}
\end{equation}
with $\kappa _{n}=\sqrt{n-3/2}$ and $n\geq 2$. These potentials  have
a quadratic dependence when $x/L\ll 1$ and a logarithmic dependence for   $%
x/L\gg 1$ such that $\mathcal{P}(x)$ $\simeq x^{-2n}$ as $\left|   x\right|
\rightarrow \infty $. It is straightforward to check that the expression
(\ref{logpot}) is consistent with the definition (\ref{hrdef}) of the  home
range $L$ for any value of $n$. In the limit of $n\rightarrow +\infty $
(\ref{logpot}) reduces to the harmonic case. We have studied, in particular,
the cases from $n=2$ to $n=8$ and have  obtained analytical expressions for the
mean square displacement. We do not display them here because they do not add
to the  understanding.
In Fig. \ref{msdcomparison} we show the cases $n=3$ and $n=8$. Already   for $%
n=8$ the saturation curve for the logarithmic and the harmonic cases  are very
close to each other. The inset of Fig. \ref{msdcomparison}  shows the  origin
of such similarities by comparing the corresponding steady state  probability
distributions $\mathcal{P}(x)$.
\begin{figure}[t]
\centering \resizebox{\columnwidth}{!}{\includegraphics{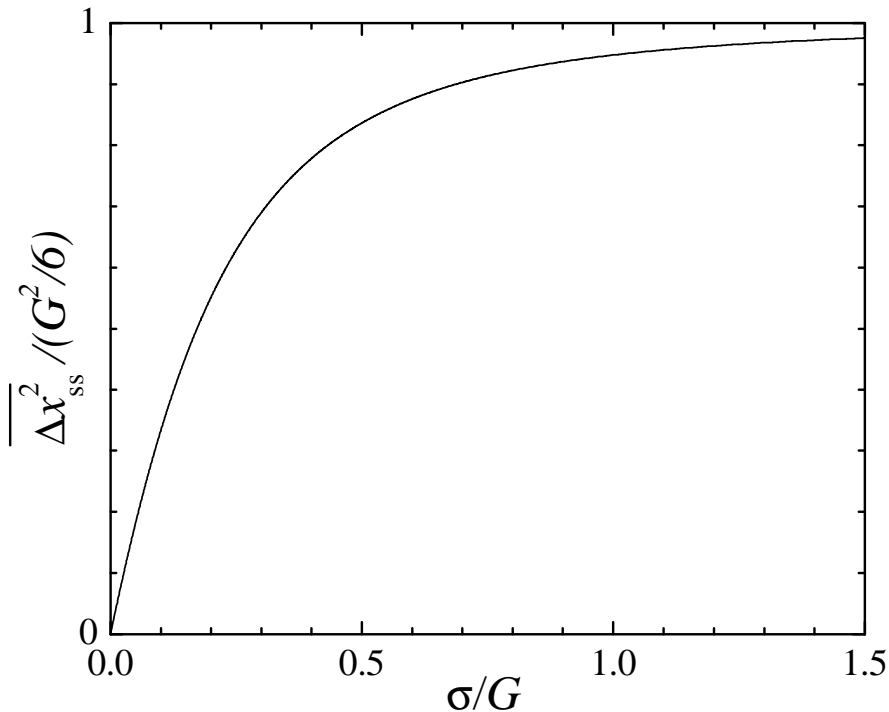}}
\caption{Mean square displacement at saturation for the potential
$U(x)=D\ln(1+(x/\sigma)^{2})$. Compared to the potentials depicted in Fig.
\ref{msdcomparison}, here $U(x)$ grows to infinity qualitatively slower. The
corresponding distribution $\mathcal{P}(x)$ does not possesses a finite second
moment. The long-tailed distribution changes drastically the behaviour of
$\overline{\Delta x_{ss}^{2}}$ when very large grid sizes are used. The growth
of the saturation curve for small $\sigma/G$ is linear and not quadratic.}
\label{msdcauchy}
\end{figure}

As mentioned above, the $\mathcal{P}(x)$ associated with the logarithmic
potentials possesses long tails. Probability distributions with long tails can
be appropriate when the motion of the animal cannot be represented by a simple
random walk. More complex types of walk may occur if the walker awaits for very
long times between jumps, or if the jumps are of very large distance.
Long-tailed $\mathcal{P}(x)$ are characterized by the feature that certain
moments of the distribution become infinite. If all the moments beyond the
first are infinite, Eq. (\ref{hrdef}) is no longer applicable for defining the
home range width. A qualitatively different behaviour is expected for large
values of $G$. To illustrate this situation, we consider the potential
\begin{equation}
U(x)=D\ln(1+\frac{x^{2}}{\sigma^{2}})
\end{equation}
whose corresponding $\mathcal{P}(x)$ is the Cauchy
distribution~\citep{lemonsbook} $\mathcal{P}(x)=(1+(x/\sigma)^{2})^{-1}/\pi$.
Also for this case the mean square displacement can be obtained analytically
and it is given by
\begin{equation}
\frac{\overline{\Delta x_{ss}^{2}}}{G^{2}/6}
=6\left[ \frac{\xi}{\tan ^{-1}\left(\frac{1}{2\xi}\right)
-\xi\ln \left( 1+\frac{1}{(2\xi)^{2}}\right)}-
4\xi^{2}\right],
\label{msdcauchyexp}
\end{equation}
where $\xi=\sigma/G$. As $\xi\rightarrow 0$, $\overline{\Delta
x_{ss}^{2}}/(G^{2}/6)\simeq (12/\pi)\sigma/G$ in Eq. (\ref{msdcauchyexp}). A
linear growth of the saturation curve emerges, as depicted in Fig.
\ref{msdcauchy}. Notice that $L$, as defined by Eq. (\ref{hrdef}), does not
exist for  this potential. The home range is therefore defined as the
characteristic length  $\sigma$, the ratio $\xi$ being the counterpart of
$\zeta$ of (\ref{zetadef}). This different qualitative behaviour with respect
to the previously analyzed cases could be exploited for determining the
characteristics of the animal  walks by making various measurements with large
grid size. A sufficient number of these measurements would allow one to discern
if the saturation curve is growing quadratically (as in the examples of  Fig.
\ref{msdcomparison}) or linearly (as in Fig. \ref{msdcauchy}).

The examples of this section illustrate that the choice of $U(x)$ in any
application of the present theory should be assessed in each case, based, for
example, on a priori  knowledge of the specific animal behavior. The different
potentials shown here give an overview of the possible qualitative behaviours
of the mean square displacement at saturation.

\section{Inhomogeneous distribution of home ranges: the case of a periodic
arrangement}

\label{sec-over}

The results obtained in Sec.~\ref{sec-msd}, from Eq.~(\ref{msdwithg})  onward,
and the examples developed in Sec.~\ref{sec-pot}, assume a continuous  and
homogeneous distribution of burrow location $x_c$. A more realistic situation
invokes the home ranges arranged in a  non-continuous manner, the centers of
adjacent ranges (the burrow locations) separated by  some characteristic
distance $a$. In this section we show an example of how this feature may be
incorporated in our analysis and and how, in principle, $a$ can be deduced from
mark-recapture  measurements.

\begin{figure}[t]
\centering \resizebox{5.5in}{!}{\includegraphics{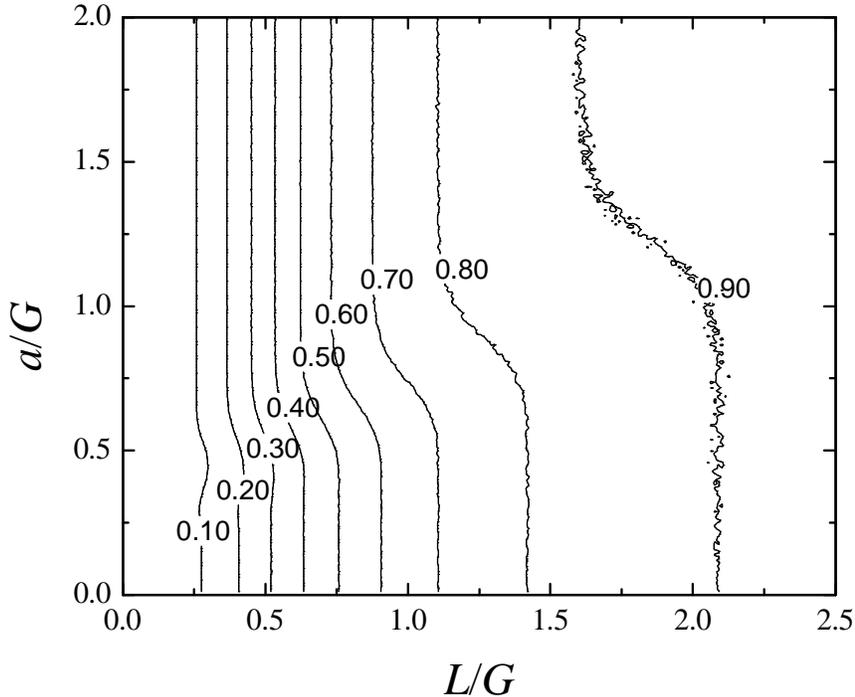}}
\caption{Contour plot of the normalized mean square displacement,
$\overline{\Delta x^{2}_{ss}}/(G^{2}/6)$, as a function of the  normalized home
range size, $L/G$ and the normalized inter-home distance, $a/G$. The  lines are
the result of simulations of the harmonic model (Gaussian occupation of
space). The fluctuations in the lines are an artifact of the construction of
the contours from discrete simulations.}
\label{contour}
\end{figure}

Let us suppose for simplicity that the home ranges are distributed in a
periodic array, with $a$ being the distance between nearest neighbors. The mean
square  displacement, measured within a window of linear size $G$, is now a
function of $a$  and $L$, besides depending parametrically on $G$. As in
Section \ref{sec-msd}, $G$ can be used to rescale the two coordinates  $a$ and
$L$. The function  $\overline{\Delta x^{2}_{ss}}/(G^{2}/6)=f(L/G,a/G)$ is
universal and does not depend on  the  size
of the observation window. We show a contour plot of this function in   Fig.~%
\ref{contour}, as calculated by numerical simulation of the harmonic   model
(Gaussian probability distributions). At $a=0$, the shape of the surface
coincides with the curve calculated in Eq.~(\ref{msdparabola}). The   contours
of equal mean square displacement are nearly vertical lines in this   plot,
indicating that the dependence on the inter-home distance $a$ is very   weak
(in particular for small values of the normalized mean square  displacement),
except for a region well defined in $a$, where the  contours shift from one
value of $L$ to another. This feature will certainly be of  relevance if an
experiment is designed in order to measure both $a$ and $L$, since  the
uncertainty on $a$  will tend to be large.

\begin{figure}[t]
\centering   \resizebox{5in}{!}{\includegraphics{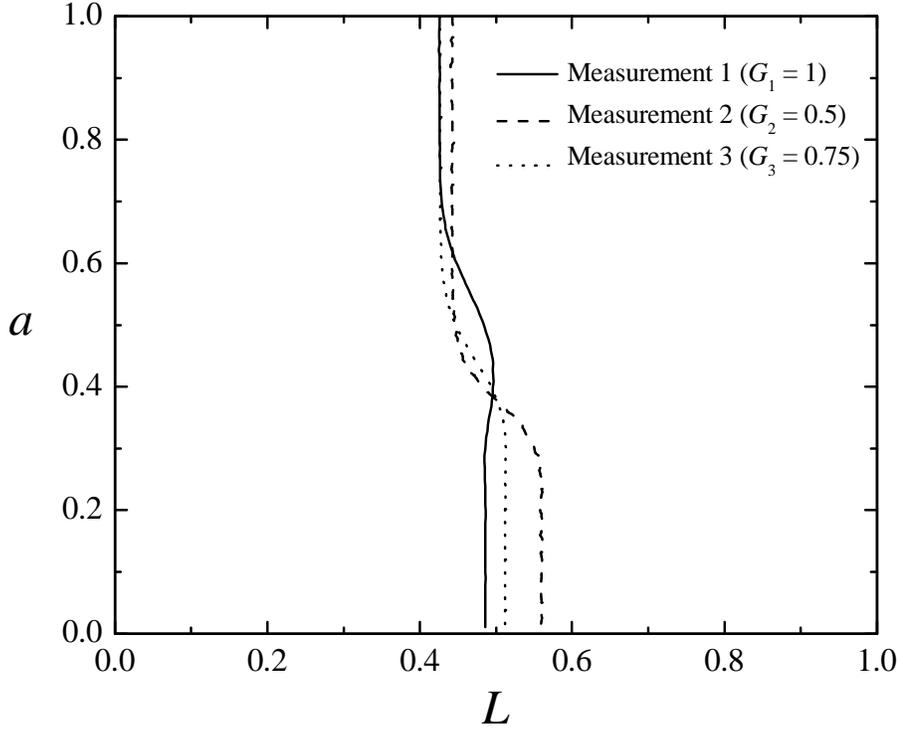}}
\caption{Graphical explanation of the procedure to determine home range
overlap via a hypothetical measurement. Both $L$ and $a$ can be determined by
taking three measurements of the same  population, using three  windows, of
sizes $G_1$, $G_2$ and $G_3$. The measured  mean square displacements  are
supposed to be $\overline{\Delta x^{2}_1}= 0.045$, $\overline{\Delta  x^{2}_2}=
0.029$, and $\overline{\Delta x^{2}_3}=  0.041$. The corresponding contours
intersect at $L=0.5$, $a=0.375$, providing  these values as a result.  Both $a$
and $L$ are displayed in units of $G$  (arbitrary linear units).}
\label{contour-g1g2g3}
\end{figure}
In general, given that the function $\overline{\Delta x^{2}_{ss}}$ is
nonlinear in both its variables $a/G$ and $L/G$, two or more  measurements are
necessary to determine the home range size and the inter-home  distance. In
Fig.~\ref{contour-g1g2g3} we show a hypothetical situation in which  three
measurements are supposed to be taken on the same population, using  three
windows sizes, $G_1=1$, $G_2=G_1/2$ and $G_3=3G_1/4$. The results of the
measurement  are curves of constant $\overline{\Delta x^{2}_{ss}}$ in the plane
$(L,a)$. With three measurements, $\overline{\Delta x^{2}_1}$,
$\overline{\Delta   x^{2}_2}$ and $\overline{\Delta x^{2}_3}$, three curves are
obtained, and the  model predicts:
\begin{eqnarray}
\overline{\Delta x^{2}_1}/(G_1^{2}/6) &=& f(L/G_1,a/G_1),  \nonumber \\
\overline{\Delta x^{2}_2}/(G_2^{2}/6) &=& f(L/G_2,a/G_2),  \nonumber \\
\overline{\Delta x^{2}_3}/(G_3^{2}/6) &=& f(L/G_3,a/G_3).  \label{array}
\end{eqnarray}
This is a system of three (nonlinear) equations with two unknowns, $L$   and $a
$, and its solution can be found as the intersection of three  curves.  These
curves are shown in Fig.~\ref{contour-g1g2g3}, displaying an  intersection very
near $L=0.5$, $a=0.375$ (within the accuracy of the fluctuations of the
contours). The curves were obtained from the  normalized function,  shown in
Fig. \ref{contour}, using the appropriate contours. As mentioned above,  the
weak dependence on $a$ may hinder its determination by the present  method. It
is clear that the choice of the appropriate values of the window sizes  is
critical to obtain the best results. This must be done specifically for  each
situation, with the help of an informed guess of the range where both $a$  and
$L$ lie. Regardless of this practical difficulty,  the procedure we describe
provides a method for an immediate measurement of an important quantity that is
hard to  obtain by other  means. Additionally, if the population is not well
characterized  by a typical inter-home distance, as we suppose here (for
example, if the  inter-home distance is bimodal due to gender differences or
other polymorphism),  the  model can be immediately modified to incorporate
those features.

\begin{figure}[t]
\centering \resizebox{\columnwidth}{!}{\includegraphics{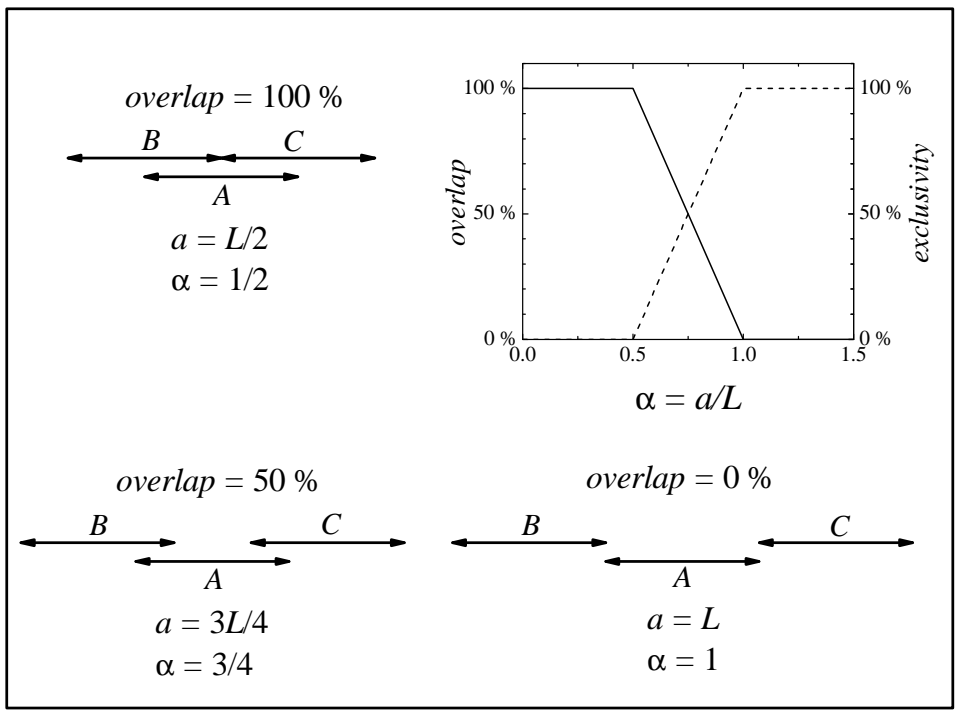}}
\caption{Illustration of home range overlap situations, and the   variables
characterizing them. The schematic diagram shows three cases: 100\%,  50\%, and
0\% overlap, with arrows representing the extent of the home range of three
neighbouring animals, $A$, $B$ and $C$. The plot shows the relation   between
the overlap, the exclusivity of space use, and the normalized inter-home
distance $\alpha=a/L$.}
\label{overlap}
\end{figure}

Moreover, the inter-home distance is closely related to the overlap of   home
ranges (or to the exclusivity of space use). See Fig. \ref{overlap} for an
illustration of three typical situations. The  home ranges of three
neighbouring animals, $A$, $B$, and $C$, are displayed as arrows, schematically
representing the extent of the area occupied by  95\% of the norm of $\mathcal
P(x)$, as usually defined \citep{worton87}. When  $a=L/2$, the first
neighbours, $B$ and $C$, of $A$ have their home ranges  situated at the border
of $A$'s home range. Then, $A$ does not have exclusive use  of any part of its
own home range (overlap equal 100\%). On the other extreme,  when $a=L$, the
home ranges of $B$ and $C$ are completely outside $A$'s. In consequence, the
exclusivity of $A$ is 100\% (home range overlap 0\%).  This value of
exclusivity, certainly, is maintained for any value of $a>L$.  An intermediate
situation, in which the exclusivity of animal $A$ is equal  to 50\% (as well as
its overlap with the neighbours) is also shown.

The exclusivity of space use has recently been found to obey an   allometric
scaling relation with the animal mass by~\citet{jetz04}. The present  theory
provides a method to determine both the home range size and the home   range
overlap, and thus to verify the scaling of these and related
magnitudes~\citep{westbrown97,banavar99}.

\section{Comparison with convex polygon calculations}

\label{sec-poly}

Home range sizes have been often deduced from the measurement of the so-called
minimum convex polygon of an animal  position. Although this procedure suffers
from a number of drawbacks  \citep{worton87}, it is used rather widely. Among
its drawbacks it is  easy to recognize at least a logical and a methodological
one. The surveyed perimeter provides no information about the use  of space
inside it, effectively encompassing areas that may be  inaccessible to the
animal,  or potentially huge areas of very low frequency of  utilization. The
methodological one is the fact that the measured area converges very  slowly to
the actual home range, and as such the observation of a few tens of  positions
provides a very bad estimation. Both kinds of flaws have been recognized in the
literature before (see  \citet{worton87} and  references therein), and
suggestions have been made to compensate for them. These proposals, such as
discarding some fraction of the extreme positions  from a set of observations
to compensate for the first, or to  join the perimetral  points in a fashion
different from the minimum convex polygon, are surely  arbitrary. Furthermore,
methodologically, they are obviously subject to  uncontrollable errors.

In the following we illustrate how the calculation of the mean square
displacement we have given in the present paper provides a rapidly converging
measurement of the home range size. This is an additional advantage when
displacements of animals belonging to a population are more readily accessible
than  repeated measurements of the position of individual  animals. For the
sake of the illustration, we will consider that the probability  of space use
of an  animal is a bivariate Gaussian distribution of variance  $\sigma$:
\begin{equation}
\mathcal{P}(x,y) =
\frac{1}{2\pi\sigma^{2}}e^{-\frac{x^{2}+y^{2}}{2\sigma^{2}}}.
\label{gaussian2d}
\end{equation}
We consider this symmetric distribution for simplicity, but the   discussion
applies equally to a distribution with anisotropic $\sigma$.  Moreover,  the
conclusions are equivalent for more general distributions, including  those
with a finite cut-off.

\begin{figure}[b]
\centering \resizebox{\columnwidth}{!}{\includegraphics{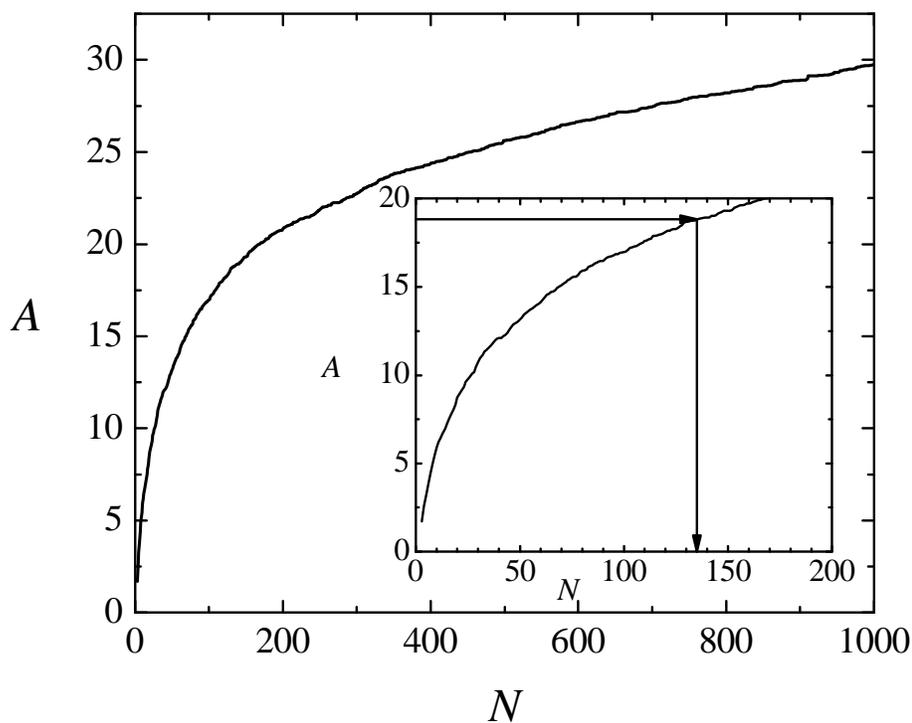}}
\caption{Area $A$ enclosed by the minimum convex polygon corresponding  to a
set of random points with Gaussian distribution in the plane, as a function of
$N$,the number of points of the set. The line shows the average of 50
independent realizations. The inset shows a detail of the same function, with
arrows showing the necessary number of points to obtain a good measurement of
the home range, defined as the 95\% of the space  occupation which, for
$\sigma=1$, and $A\approx 18.8$, is $N=135$ points. The  variance $\sigma$ has
arbitrary units of length, and $A$ those of length squared.}
\label{polygon}
\end{figure}

If we define the home range of the animal whose space use distribution  is
described by (\ref{gaussian2d}) as the area $A$ that contains the 95\% of
$\mathcal{P}$~\citep{worton87}, a simple integration gives
\begin{equation}
A = \pi R^{2} = 2\pi\sigma^{2}(-\ln 0.05)\approx 18.8\sigma^{2}.
\label{hrsize}
\end{equation}
This the quantity that we intend to measure by both methods. In Fig.
\ref{polygon} we show the area of the minimum convex polygon   defined by a set
of $N$ points drawn at random with a bivariate Gaussian  distribution of
$\sigma=1$. Observe, firstly, that the area grows unboundedly as the  number of
observations grows, since $\mathcal{P}$ is unbounded. More relevant from the
practical point of view is the fact that the growth is very slowly, and that
the area $A$ (obtained with 95\% rule  and marked in Fig. \ref{polygon} with an
arrow) is achieved after the observation contains, on average, 135  points (see
the inset of Fig. \ref{polygon}).

\begin{figure}[b]
\centering \resizebox{\columnwidth}{!}{\includegraphics{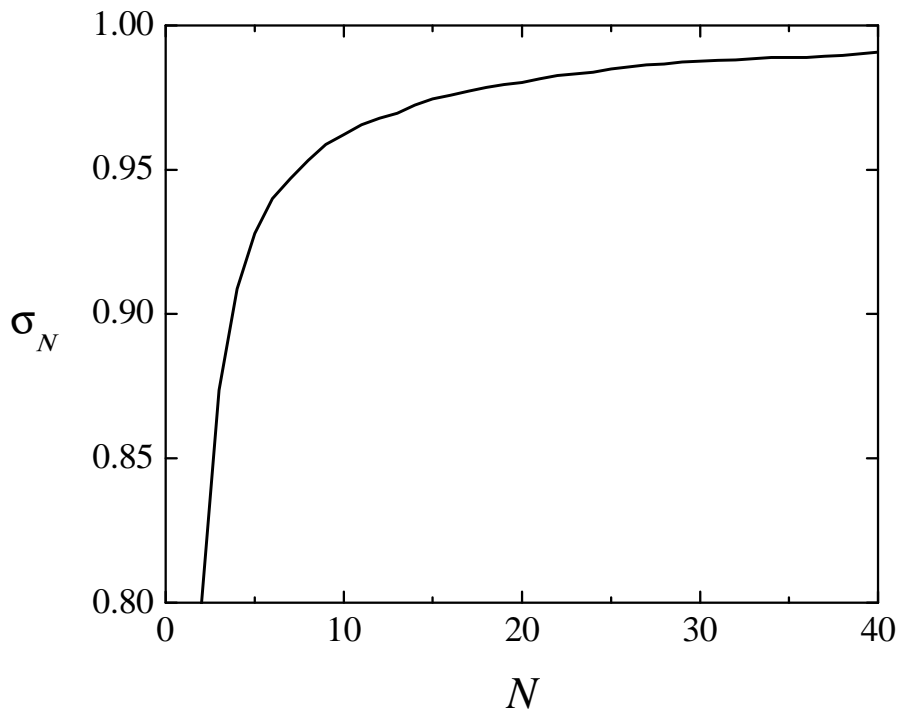}}
\caption{Measurement of the variance of a bivariate Gaussian   distribution ($%
\sigma = 1$), as a function of the number of displacements used for the
estimate. The curve shows an average based on $104$ realizations.}
\label{sigma}
\end{figure}

Now consider that one wants to determine $A$ by measuring \emph{displacements},
instead of positions. The  relevant quantity to be estimated is the variance
$\sigma$, immediately derived   from the home range size $L$, accessible
through our present theory  explained in  Sec. \ref{sec-msd}. Expression
(\ref{hrsize}) gives then the home range  area.  Figure \ref{sigma} shows that
the measurement of the variance, $\sigma_N$, for   a finite set of $N$
observations, converges very rapidly to the actual  value (which is 1 in this
case), when the number of observations is increased.  Indeed, with just 10
observations of the displacement the variance can  be estimated  with an
accuracy greater than 95\%. Ten displacements correspond to just 11  positions
of a single animal, if taken at intervals long enough that  they are
uncorrelated, or 10 displacements of different animals for which an average
distribution would be found.

In summary, the mean square displacement provides a faster convergence  to the
area of the home range than the construction of the convex polygon. In
addition, the estimation of the distribution $\mathcal{P}$ (unimodal  in this
case, but easily generalizable) by that method provides information about the
use of the home range, which is inaccessible to the convex polygon procedure.

\section{Conclusions}
\label{sec-conc}

The determination of home range dimensions and spatial overlap of two
neighbouring home ranges from field observation is a subject of great interest
for the understanding of animal motion. We have provided here a general theory
to extract such demographic  parameters on the basis of the measurement of
displacements of individual animals  in a population.

The most common techniques for gathering information of home range size employ
trapping of animal and radiotelemetry observations. Our theory in the present
paper has been constructed with the specific  goal of interpreting data
obtained from the former type of observation, i.e., mark-recapture experiments,
in which the sampling area is finite.

The motion of an animal inside its own home range has been modeled by a
Fokker-Planck equation (\ref{FPeq}), i.e. by diffusion in a confining potential
$U(x)$. While the equations considered are, for simplicity, 1-dimensional,
extension to higher dimensions is straightforward and unnecessary for practical
purposes.

Even though the precise determination of the home range size $L$  depends on
the choice of the potential $U(x)$, the general sigmoid shape of the saturation
curve in our theory indicates that the difference in the results is not
substantial if the window   size is chosen such that $L<G$. Eventually, the
choice of the right potential is to be determined for each given case, on the
basis of biological  information of the  animal population under study. We have
shown that, for those situations in which the second moment of the distribution
$\mathcal{P}(x)=\exp(-U(x)/D)$ is not finite, the saturation curve for large
$G$ grows linearly with $L$ rather than quadratically as in more conventional
potentials (box or harmonic). Such cases may arise when the animal motion is
not  simple but involve a more complicated random walk such as a Levi walk or
flight. This means that our theory of mark-recapture observation may be used,
in principle, to determine whether the animal population is performing a
Gaussian random walk or a more  complicated walk. By measuring the mean square
displacement at saturation with different  values of the probe length $G$
(sufficiently lager than $L$) it might be  possible to determine if the
saturation curve grows quadratically or linearly.

The obtained parameters that characterize the average use of space, when
obtained via mark-recapture observations and their interpretation  with our
present theory, converge rapidly to the expected values. We have shown in Sec.
\ref{sec-poly} that this is not the case in the  application of the traditional
minimum convex polygon method. We thus suggest that displacement measurements
of an animal population  should be considered as the most appropriate way for
determining home range  dimensions if radiotelemetry methods are not available.

The other important demographic parameter that can be extracted from
mark-recapture measurements is the inter-home characteristic distance  of the
animals, called $a$ in the present paper. Such a length is simply related to
the mean overlap (or the mean exclusivity of space usage). We have outlined a
procedure to extract this parameter quantitatively from mark-recapture
observations and provided a general way to verify directly from mark-recapture
experiments the scaling of home range overlap as function of body mass.

\begin{ack}
It is a pleasure to thank Marcelo Kuperman and Ignacio Peixoto for  fruitful
discussions. This work was supported in part by the NSF under grant no.
INT-0336343, by NSF/NIH Ecology of Infectious Diseases under grant no.
EF-0326757 and by DARPA under grant no. DARPA-N00014-03-1-0900. G.  Abramson
acknowledges partial funding by CONICET (PEI 6482), and by  Fundaci\'{o}n
Antorchas.
\end{ack}

\end{document}